\documentstyle[psfig]{article}

\def\abstract#1{\vskip 7mm 
        \begin{center}{\large Abstract}\par \smallskip
                \begin{minipage}[c]{12cm}
                        \small #1
                \end{minipage}
        \end{center}
}
\def\title#1{\begin{center}{\Large\bf #1}\end{center}}
\def\author#1{\vskip 5mm \begin{center}{#1}\end{center}}
\def\address#1{\begin{center}{\it #1}\end{center}}

\newcommand{\k}{{\mathbf{k}}}

\newcommand{\ti}{\textit}
\newcommand{\f}{\frac}

\newcommand{\bb}{\bibitem}
\newcommand{\BF}{\begin{figure}\begin{center}}
\newcommand{\EF}{\end{center}\end{figure}}
\newcommand{\BE}{\begin{equation}}
\newcommand{\EE}{\end{equation}}
\newcommand{\BEA}{\begin{eqnarray}}
\newcommand{\EEA}{\end{eqnarray}}
\makeatletter
\@ifundefined{lesssim}{}{}
\@ifundefined{gtrsim}{}{}
\def\vereq#1#2{\lower3pt\vbox{\baselineskip1.5pt \lineskip1.5pt
\ialign{$\m@th#1\hfill##\hfil$\crcr#2\crcr\sim\crcr}}}
\makeatother

\begin{document}

\title{
  Cosmic Microwave Background in Closed Multiply Connected Universes
  }
\author{
  Kaiki Taro Inoue,\footnote{E-mail:tinoue@yukawa.kyoto-u.ac.jp}
}
\address{
  Yukawa Institute for Theoretical Physics, Kyoto University, \\ 
  Kyoto 606--8502, Japan
}
\abstract{We have investigated the cosmic microwave background (CMB) 
anisotropy in closed multiply connected universes (flat and hyperbolic) 
with low matter density.
We show that the 
COBE constraints on these low matter density models with non-trivial
topology are less stringent since 
a large amount of CMB anisotropy on 
large angular scales can be produced 
due to the decay of the gravitational potential at late time. }

\section{Introduction}
For a long time, cosmologists have assumed the simply connectivity
of the spatial hypersurface of the universe. If it is the case, 
the topology of closed 3-spaces is limited to that of a 3-sphere 
if Poincar\'{e}'s conjecture is correct.  
However, there is no particular reason for assuming
the simply connectivity since the Einstein equations do not
specify the boundary conditions. 
If we allow the spatial 
hypersurface being multiply connected then the spatial geometry 
of closed models can be flat or hyperbolic as well.
We should be able to
observe the imprint of the ``finiteness'' of the spatial geometry if
it is multiply connected on scales of the order of the 
particle horizon or less, in other words, if we live in a 
``small universe''. 
\\
\indent
However, it has been claimed by several authors that the ``small
universes'' have already been ruled out observationally. 
For a flat 3-torus model without the cosmological constant,
the COBE data constrains the topological identification scale $L$ (the 
minimum length of periodic geodesics) to be $L\ge 0.4\times 2 R_\ast$
where $R_\ast$ is the comoving radius of the last scattering surface
\cite{Sokolov,Starobinsky,Stevens,Oliveira1}.
The suppression of the fluctuations on scales beyond $L$ leads to 
a decrease of the angular power spectrum $C_l$ of the CMB
temperature fluctuations on large angular scales.
\\
\indent
In contrast, for low matter density models, the constraint could be
considerably less stringent
since a bulk of large-angle CMB fluctuations 
can be produced by the so-called (late) integrated Sachs-Wolfe effect
\cite{Cornish2} which is the gravitational blueshift
effect of the free streaming photons caused by 
the decay of the gravitational potential\cite{HSS}. 
If the background geometry is either flat or hyperbolic,
then the angular size of a fluctuation becomes small as it
approaches to the observation point. Although fluctuations beyond the
size of the fundamental domain are suppressed, no significant
suppression in large-angle power occurs if they are produced at place
well after the last scattering. In other words, large-angle fluctuations
can be generated when the fluctuations enter the 
topological identification scale at late time. 
\section{Closed Flat Models}
Let us first consider closed models in which the spatial geometry is
represented as a flat 3-torus obtained by gluing 
the opposite faces of a cube with sides $L$ 
by three translations. Then the wave numbers
of the square-integrable eigenmodes of the Laplacian are restricted to
the discrete values $k_i=2 \pi n_i/L$, ($i=1,2,3$) where $n_i$'s run
over all integers. Assuming adiabatic initial perturbation, 
the angular power spectrum is written as
\BEA
C_l&=&\sum_{\k\ne 0} \f{8 \pi^3 {\cal{P}}_\Phi(k)F_{kl}^2}{k^3 L^3},
\nonumber
\\
F_{kl}&=&\f{1}{3}
\Phi(\eta_\ast) j_l(k(\eta_o\!-\!\eta_\ast))
\!+2 \int_{\eta_\ast}^
{\eta_o}\!\!\!\!\!\!d \eta\, 
\f{d\Phi}{d \eta}j_l(k(\eta_o\!-\!\eta)),
\label{eq:psT3}
\EEA
where $\eta_\ast$ and $\eta_0$ correspond to the last scattering 
and the present conformal time, ${\cal{P}}_\Phi(k)$ 
is the initial power spectrum for the
Newtonian curvature perturbation $\Phi$
and $k\equiv\sqrt{k_1^2+k_2^2+k_3^2}$. From now on we assume
the scale-invariant 
Harrison-Zel'dovich spectrum (${\cal{P}}_\Phi(k)
\!=\!const.$) as the initial power spectrum. 
The angular scale which gives the suppression scale $l_{cut}$ is
determined by the lowest eigenmode 
on the last scattering surface.
The oscillation scale for $l\!<\!l_{cut}$ is also 
determined by the first eigenmode corresponding to 
the contribution from ordinary Sachs-Wolfe effect and the
integrated Sachs-Wolfe effect. On smaller angular scales 
$l>l_{cut}$, each peak in the power corresponds to the
fluctuation scale of the second and higher eigenmodes 
at the last scattering. 
This behavior is analogous to the acoustic oscillation where the
oscillation scale is determined by the sound horizon at the 
last scattering.
\BF
\centerline{\psfig{figure=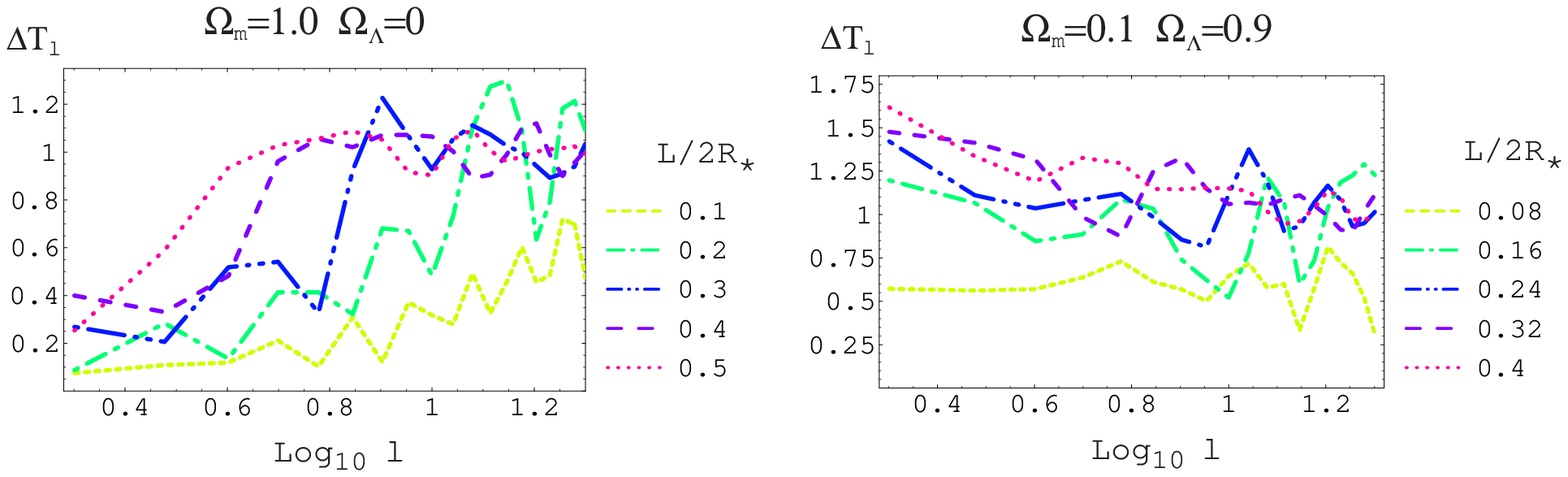,width=17cm}}
\caption{Suppression in large-angle power for 3-torus models with or
without the cosmological constant. 
The significant large-angle suppression 
in $\Delta T_l\equiv \sqrt{l(l+1)C_l/(2
\pi)}$ (all the plotted values are 
normalized by $\Delta T_{20}$ with infinite volume) 
occurs for the ``standard'' 3-torus model 
$(\Omega_m,\Omega_\Lambda)\!=\!(1.0,0)$
at $l\!<\!l_{cut}\!\sim\! 2\pi R_\ast/L-1$ 
while such prominent suppression is
\ti{not} observed for the 3-torus model with 
($\Omega_m, \Omega_\Lambda$)=(0.1,0.9).}
\label{fig:TLF}
\EF
As shown in 
figure 1, the angular power for a model with
$(\Omega_m,\Omega_\Lambda)\!=\!(0.1,0.9)$ is jagged in $l$ but 
strong suppression is not observed for even very small models.
Surprisingly, in low matter density models,  
the slight excess power due to the integrated Sachs-Wolfe 
effect is cancelled out
by the moderate suppression owing to the mode-cutoff which leads to
a nearly flat spectrum. However, as observed in the ``standard'' 3-torus 
model, the power spectra have prominent oscillating features. 
\\
\indent
We have carried out 
Bayesian analyses using the COBE-DMR 4-year data and obtained
the constraints in the size of the fundamental domain 
$L\ge 1.6H_0^{-1}$ and $L\ge2.2 H_0^{-1}$ 
for the ``standard'' 3-torus model $(\Omega_m,\Omega_\Lambda)=(1.0,0)$ 
and the 3-torus model with low matter density 
$(\Omega_m,\Omega_\Lambda)=(0.1,0.9)$,
respectively. The maximum number $N$ of images of the cell  
within the observable region at present is 8 and 49 for the
former and the latter, respectively\cite{Inoue7}. 
\section{Closed Hyperbolic Models}
Next we consider closed hyperbolic models with small volume.
The interesting property of 
closed (compact) hyperbolic manifolds is the existence of the lower bound
for the volume $V\!>\!0.16668...$ (in unit of cube of curvature radius).
If the creation of the universe with smaller
volume is more likely then it gives the reason why the topological
identification scale is comparable to the present horizon scale. 
For flat models, there is no reason for the ``coincidence'' in the scale
since one can choose the volume arbitrarily. 
The smallest known one is called the Weeks manifold with volume 
0.94 (in unit of cube of curvature radius).
Today, the number of known examples of closed hyperbolic manifolds is
more than 10000 which have been stored in a computer program ``SnapPea''
by J. Weeks\cite{SnapPea}. 
\BF
\centerline{\psfig{figure=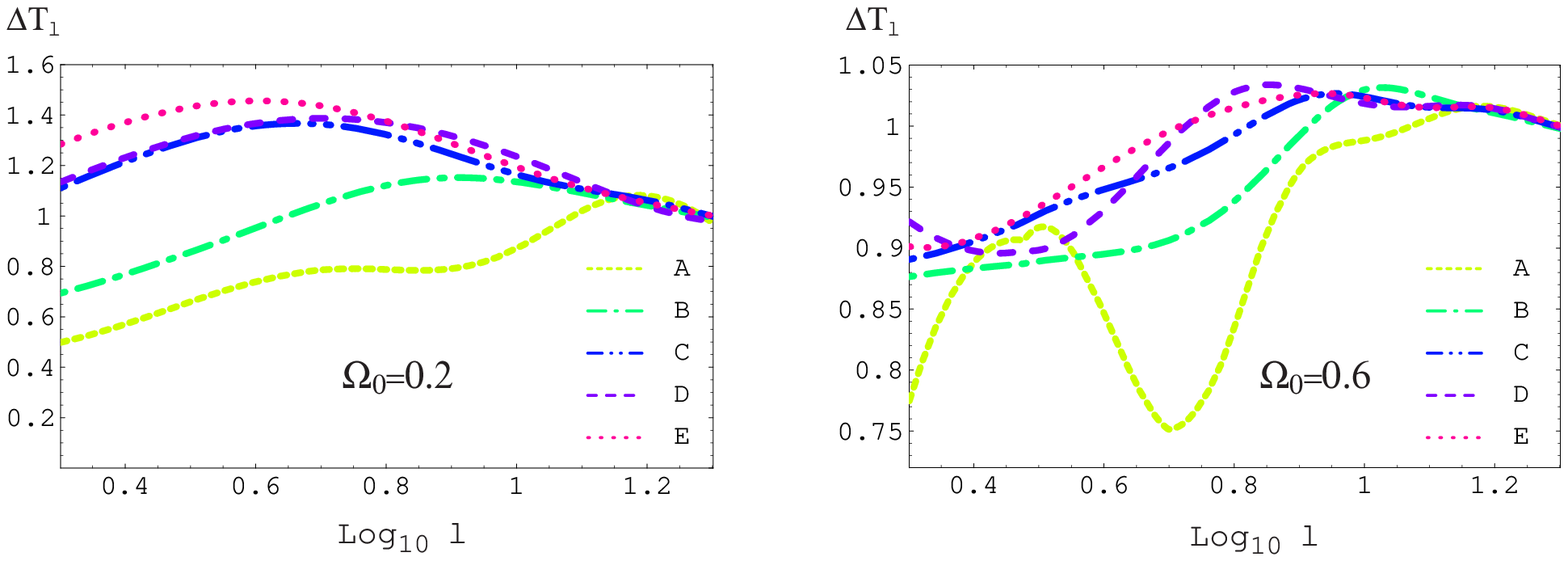,width=17cm}}
\caption{Suppression in large-angle power 
$\Delta T_l\equiv \sqrt{l(l+1)C_l/(2
\pi)}$ for five closed hyperbolic 
models (name,volume)=A:(m003(3,-1),0.94), B:(m010(-1,3),1.9),
C:(m082(-2,3),2.9), D:(m288(-5,1),3.9) and E:(s873(-4,1), 4.9).}
\label{fig:TLCHM5}
\EF
\\
\indent
In fact, we have found that the suppression in the angular power
due to the mode-cutoff for closed hyperbolic models is also very
weak (figure 2).  As is the case for flat 3-torus models, the 
suppression scale $l_{cut}$ is determined by the lowest eigenmode 
on the last scattering surface. 
Beyond the scale $l_{cut}$ the ordinary Sachs-Wolfe 
contribution is strongly suppressed as 
in compact flat models without the $\Lambda$ term. 
The weak suppressions imply the significant 
contribution from the integrated Sachs-Wolfe effect on 
large-angle scales $l>l_{cut}$. Because the curvature dominant epoch
comes earlier in time than the $\Lambda$ dominant epoch, the contribution 
from the integrated Sachs-Wolfe effect is much greater than that 
for flat $\Lambda$ models.  
Below the scale $l_{cut}$, a considerable
amount of contribution comes from the second lowest eigenmode. 
As the number of modes which contribute to $C_l$ grows,
$C_l$ converges to the value for the infinite counterpart. 
The obtained constraint for the Weeks model (volume=0.94)
is $\Omega_0\ge0.1$ and the maximum expected 
number of copies of the fundamental
domain inside the present observable region is approximately 1560.
Note that the obtained result agrees with other recent
works\cite{Aurich1,CS,Aurich2}.  
Furthermore, for some closed hyperbolic models, it is found that 
the computed angular powers give a much better fit to the COBE 
data since the quadrapole is very low. 
\\
\indent
However, one might argue that the constraint using only the power spectrum
is not sufficient since it contains only isotropic information for
2-point correlations\cite{Bond}. In fact there is a gap between the 
likelihood using only the power spectrum and that using full
covariance elements for the 3-torus models\cite{Inoue7}. For 
globally anisotropic models, the fluctuations 
are anisotropic Gaussian for a given axis but are non-Gaussian if
the likelihoods are marginalized over the axis. For locally closed 
Friedmann-Robertson-Walker models, the skewness is zero but the
kurtosis is non-zero assuming that the initial fluctuations are Gaussian.
Another feature is the correlation between the expansion coefficient 
$a_{l m}$'s of the temperature fluctuations in the sky which are
independent random numbers for the standard infinite counterparts
in which the initial fluctuations are homogeneous and
isotropic Gaussian. In the case of flat 3-torus models, $a_{l m}$'s 
are written in terms of spherical harmonics $Y_{l m}(\hat \k)$
which correspond to the expansion coefficients of the eigenmodes in the 
3-torus in terms of the eigenmodes of the 3-dimensional
Euclidean space. Therefore, if the number of the term in the sum
that gives $a_{l m}$ is small then  $a_{l m}$'s are no longer
independent. This is the main reason for the gap in the likelihoods 
for the 3-torus models. 
In contrast, for closed hyperbolic models, the expansion coefficients
of the eigenmodes are well approximated by Gaussian pseudo random
numbers \cite{Inoue1,Inoue3}. 
Therefore for homogeneous ensembles, $a_{l m}$ can
be described as independent ``random'' numbers although 
they are non Gaussian (since they are written in terms of a sum of
products of two independent Gaussian numbers). 
This implies that the value of the likelihood is 
highly \ti{dependent on the place} of the observer in the manifold.
In order to constrain closed hyperbolic models which are globally
inhomogeneous, one should calculate the likelihood everywhere 
in the manifold. If one marginalizes the likelihood over
the orientation (axis) and the position of the observer then
 $a_{l m}$'s become independent non-Gaussian random numbers.
Thus we expect that the constraints using only the power spectrum
give better estimates for closed hyperbolic models than closed
flat models in which the correlations in $a_{l m}$'s are prominent.  
\section{Conclusion}
We have investigated the CMB anisotropy in 
closed multiply connected universes (flat and hyperbolic) 
with low matter density. We have seen that the COBE constraints
for these models are less stringent compared with the simplest ``standard'' 
3-torus model ($\Omega_m\!=\!1.0$). On the other hand 
recent observations of distant supernova 
Ia \cite{Perlmutter,Riess} and of the CMB on small
angular scales \cite{Boomerang,MAXIMA} imply that 
our universe is nearly flat with the cosmological constant
(or ``quintessence'', X-matter etc.) which dominates the present
universe. It seems that the closed hyperbolic models are ruled out
but it is still not conclusive. 
If one includes the 
cosmological constant for a fixed curvature radius,
the comoving radius of the last scattering surface in 
unit of curvature radius becomes large. Therefore the observable
effects of the non-trivial topology become much prominent.  
For instance, the number $N$ of copies of the fundamental domains inside
the observable region at present is approximately 28
for the Weeks model with $\Omega_\Lambda\!=\!0.6$ and $\Omega_m\!=0.2\!$ 
whereas $N\!=\!4$ if $\Omega_\Lambda\!=\!0$ and $\Omega_m\!=0.8\!$.
If one allows hyperbolic orbifold models then the number can be increased as
much as $N=52$ for $\Omega_\Lambda\!=\!0.75$ and $\Omega_m\!=\!0.2$ 
since the volume of the smallest orbifold (arithmetic) is very small
(=0.039). At least at the classical level it seems that 
there is no reason to exclude any orbifold models although they have 
singular points where the curvature diverges.

\end{document}